\documentclass[pre,twocolumn,aps,superscriptaddress,showpacs,floatfix]{revtex4}
\usepackage{graphicx}
\usepackage{amsmath}
\begin{document}

\title{Brownian motors: current fluctuations and rectification efficiency}

\author{L. Machura}
\affiliation{Institute of Physics, University of Augsburg,
Universit\"atsstrasse 1, D-86135 Augsburg, Germany}
\affiliation{Institute of Physics, University of Silesia,
P-40-007 Katowice, Poland}
\author{M. Kostur}
\affiliation{Institute of Physics, University of Augsburg,
Universit\"atsstrasse 1, D-86135 Augsburg, Germany}
\author{P. Talkner}
\affiliation{Institute of Physics, University of Augsburg,
Universit\"atsstrasse 1, D-86135 Augsburg, Germany}
\author{J. \L uczka}
\affiliation{Institute of Physics,  University of Silesia,
 P-40-007 Katowice, Poland}
\author{F. Marchesoni}
\affiliation{Dipartimento di Fisica, Universit\`{a} di Camerino, 
I-62032 Camerino,  Italy}
\author{P. H\"anggi}
\affiliation{Institute of Physics, University of Augsburg,
Universit\"atsstrasse 1, D-86135 Augsburg, Germany}
\date{\today}
\begin{abstract}
  With this work we investigate an often neglected aspect of Brownian
  motor transport: The r\^{o}le of fluctuations of the noise-induced
  current and its consequences for the efficiency of rectifying noise.
  In doing so, we consider a Brownian inertial motor that is driven by
  an unbiased monochromatic, time-periodic force and thermal noise.
  Typically, we find that the asymptotic, time- and noise-averaged
  transport velocities are small, possessing rather broad velocity
  fluctuations. This implies a corresponding poor performance for the
  rectification power. However, for tailored profiles of the ratchet
  potential and appropriate drive parameters, we can identify a
  drastic enhancement of the rectification efficiency. This regime is
  marked by persistent, uni-directional motion of the Brownian motor
  with few back-turns, only. The corresponding asymmetric velocity
  distribution is then rather narrow, with a support that
  predominantly favors only one sign for the velocity.
\end{abstract}
\maketitle

\section{INTRODUCTION}
The channelling of particles by harvesting the thermal noise gives
rise to a diffusive transport of particles. In periodic potentials the
second law of thermodynamics implies that no net transport occurs.
The situation changes drastically, however, in the presence of
unbiased non-equilibrium noise which acts additionally on such
systems. Then, the concept of Brownian motors \cite{BM} does provide
the possibility for directed, noise-induced transport.  The phenomenon
has widespread applications in physics, chemistry and in the
biological sciences where it can be put to work for shuttling reliably
and efficiently particles on the micro-scale, or even on the
nano-scale \cite{BM}.

The vast majority of works on Brownian motors concentrate on the
behavior and the selective control of the emerging directed transport
as a function of control parameters such as the temperature $T$, an
external load $F$ (yielding the load-current characteristics), or some
other control variable. In contrast, the role of the fluctuations of
the directed current has not attracted much attention in the
literature.
A notable exception is the
first work on an inertial (rocking) ratchet \cite{Jung1996} wherein
the higher order, statistical cumulant properties of the stochastic
position variable have been explored.
Here, we fill this gap and focus in more detail on the {\it
fluctuating behavior} of the Brownian motor current.
The average
drift motion together with its fluctuation statistics are
salient features when characterizing the {\it modus
operandi} of a particular Brownian motor.

It is intuitive that the fluctuations of the drift variable do impact
the overall `` efficiency'' of the transport under consideration. The
objective of an optimal operation of a Brownian motor machine can be
formulated in a variety of ways, e.g. see in \cite{Sanchez2003}.  In
close analogy to heat-engine machines, one can define a generalized,
{\it nonequilibrium} efficiency of a Brownian motor.  There exists no
universally agreed upon definition of this notion \cite{efficiency,
  Derenyi1999,Oster, Suzuki2003} -- for a recent review on efficiency
of Brownian motors see Ref. \cite{parrondo}.  The most common
definition of the efficiency of a Brownian motor operating at
non-equilibrium is based on the ratio of the work (or power) done by
the particle against an external load and the input energy (input
power). With this working definition the load force is inevitably
included; in particular, this yields the result that the efficiency
assumes a {\it zero}\/ value when {\it no} load force is acting.

Alternative proposals for efficiencies have been proposed as well
\cite{Derenyi1999,Oster,Suzuki2003}: Some of these proposals do provide a
{\it nonzero} value for a vanishing bias force. Yet another
possibility consists in characterizing the rectification power of the
transport in terms of the so-called Peclet number, i.e. the quotient
of the drift velocity and the diffusion.  This notion has been used in
continuous state Brownian motor transport \cite{Schreier1998}, and
also for discrete motor models \cite{Freund1999}.  The concept is
related in spirit to a Fano factor  measure \cite{jung2002} of the
velocity fluctuations used, for example, to characterize molecular
motors \cite{kolomeisky2003}.

In this work, we shall follow the reasoning of Suzuki and Munakata
\cite{Suzuki2003}, in order to characterize the efficiency of
rectification in absence of external bias forces. Typically, there
occurs a competition between two mechanisms: a ``giant enhancement''
of diffusion \cite{Reimann2001,Marchesoni,Reimann2002,Schreier1998}
and an optimally large, (uni)-directional transport velocity
\cite{Lindner2001,Bier2003}.  The first perspective aims at
controlling the magnitude of the effective diffusion independently of
the temperature. It thus carries a rich potential for technological
separation devices.  The second facet attempts to achieve a maximal
``coherence'' for the transport.  Such a coherence is of relevance for
Brownian motors modelling biophysical molecular motors
\cite{molecular}.

Most of the Brownian motors and the majority of applications studied
in the prior literature operate in the so-called overdamped Brownian
motion regime. For specific applications the r\^{o}le of inertial
effects can become, however, of salient importance
\cite{Jung1996, Lindner1999, Mateos2000, Flach2000, Barbi, Son2003,
  Borromeo2002, Family, Sintes, Sengupta2004, Fleishman2004,
  Marchesoni}. The overdamped dynamics is a valid approximation for
many physical applications \cite{BM}.  It is also particularly well
suited to describe the motion of molecular motors \cite{BM,molecular}.
In other situations, however, the inertial effects cannot be
neglected. An exemplar is the diffusion of atoms on a crystal
surface \cite{Pollak1993}. There, the dynamics may be
underdamped, exhibiting long correlated hopping among binding sites.
This physics has been verified experimentally by use of scanning
tunnelling microscopy \cite{Ganz1992}, field ion microscopy
\cite{Senft1995}, or for quasi-elastic Helium atom scattering
\cite{Ellis1993}.

The inclusion of inertia adds significant complexity to the problem.
This is so, because a periodically rocked, single degree of freedom
with nonzero mass possesses a three-dimensional phase space that can
exhibit a chaotic dynamics \cite{Jung1996, Mateos2000, Flach2000}.
This is in contrast to the case of rocked, overdamped Brownian motors
\cite{Bartussek1994, Zapata1996, Luczka1995, Savelev2004,
  Heinsalu2004}.  While the chaotic dynamics of a driven-damped
particle in a symmetric periodic potential has been investigated
thoroughly during the 1980's, the (chaos)-induced, directed transport
of an asymmetric, inertial Brownian motor has been pioneered only much
later in \cite{Jung1996}.  There, it has been demonstrated that the
corresponding dynamics features a rich structure, possessing many
intriguing current-reversals.  In the deterministic case, different
asymptotic solutions can coexist, e.g. running and locked states.
Moreover, the onset of diffusive behavior due to chaotic dynamics has
been investigated in terms of the second moment of the particle
position diffusion.  This first study was followed up with more
detailed investigations \cite{Mateos2000}, where it has been shown,
for example, that the transport may reverse the direction at the
transition from a chaotic to a regular motion.  Additionally,
intermittent trajectories have been observed. In such cases the system
follows for a certain time a regular orbit, but then suddenly switches
to a sticking orbit. The resulting average flux depends on the time
which the particle spends in a particular state, and the system may
exhibit super-diffusive behavior.  In related work
\cite{Borromeo2002}, the inertial Brownian motor dynamics in a regime
of adiabatic driving has been investigated. These authors focused on
the onset of the diffusive transport as the damping coefficient
decreases.

In this work we concentrate on the connection between the directed
transport and its fluctuation characteristics. This study is of
relevance for the optimization of rectification: The directed current
should not become
swamped with the unavoidable fluctuations of the transported quantity.
Our investigation is based on
an inertial, noise-driven rocking ratchet.  We shall analyze
quantities such as: the long-time averaged velocity, its velocity
fluctuations, the fluctuations of the kinetic energy of the Brownian
motor and the efficiency for rectification.

The paper is organized as follows. In the next section we present the
Brownian motor model. In Section III, we elaborate on  the
problem of the efficiency for rectifying noise in connection with the
fluctuation behavior of the Brownian motor current.  In Section IV, we
describe our numerical findings for a generic set of parameters, while
in Section V, we elucidate the optimal working conditions for
rectification and directed transport.

\section{MODEL}

To start, we consider the motion of a classical particle of mass $m$
moving in the periodic, asymmetric ratchet potential
$V(x)$. The particle is driven by an unbiased time-periodic,
monochromatic force of  strength $A$ and an angular frequency
$\Omega$. The dynamics is additionally subjected to  thermal
noise.  The Brownian motor dynamics is thus governed by the Langevin
equation \cite{hanggi1982}
\begin{eqnarray}
 \label{eq:Lan1}
 m \ddot x + \gamma \dot x = -V'(x) + A \cos(\Omega t) + \sqrt{2\gamma
 kT} \; \xi(t),
\end{eqnarray}
where the prime denotes a differentiation with respect to the argument
of $V(x)$. The parameter $\gamma$ is the friction coefficient, $T$
denotes temperature, and $k$ is the Boltzmann constant.  The ratchet
potential $V(x)=V(x+L)$ has the period $L$ and a barrier height
$\Delta V$. Thermal fluctuations are modelled by the
zero-mean $\delta$-correlated Gaussian white noise $\xi(t)$. This noise
term obeys the Einstein relation with the noise correlation
given by $\langle \xi(t)\xi(s)\rangle = \delta(t-s)$.  We next
introduce dimensionless variables. The natural length scale is given
by the period $L$ of the ratchet potential. The dynamics possesses
several time scales.  We introduce the characteristic time $\tau_0$ as
determined formally from the Newton equation, $m\ddot x=-V'(x)$, by balancing
 the two forces $mL/\tau_0^2 = \Delta V/L$; yielding $\tau_0^2
= mL^2/\Delta V$.  The scaled variables thus read:
\begin{eqnarray}
\hat{x} = \frac{x}{L}, \qquad \hat{t} = \frac{t} {\tau_0}.
\end{eqnarray}
The dimensionless Langevin dynamics consequently assumes the form
\begin{equation}
\ddot{\hat x} + \hat{\gamma} \dot{\hat x} =- \hat{V}'(\hat{x}) + a
\cos(\omega \hat{t}) + \sqrt{2\hat{\gamma}D_0} \; \hat{\xi} (\hat{t}),
\label{NLbw}
\end{equation}
where
\begin{itemize}
\item the re-scaled friction coefficient $\hat{\gamma} = (\gamma / m)
\tau_0$ is the ratio of the two characteristic time scales, $\tau_0$
and the relaxation time scale of the velocity degree of freedom, i.e.,
$\tau_L = m/\gamma$,
\item the potential $\hat{V}(\hat{x})=V(x)/\Delta V$ assumes the
period $1$, and the barrier height equals unity; i.e. $\Delta \hat{V}=1$,
\item the drive has the re-scaled force strength $a = A L / \Delta V$
with the dimensionless angular frequency $\omega = \Omega \tau_0$,
\item the re-scaled, zero-mean Gaussian white noise forces
$\hat{\xi}(\hat{t})$ obey
$\langle\hat{\xi}(\hat{t})\hat{\xi}(\hat{s})\rangle=\delta(\hat{t}-\hat{s})$
with a re-scaled noise intensity $D_0 = kT / \Delta V$.
\end{itemize}
In the following, mostly for the sake of simplicity, we shall only use
dimensionless variables and shall omit the ``hat''-notation  in all
quantities.  For the asymmetric ratchet potential $V(x)$ we consider a linear
superposition of at least two (or even three) spatial harmonics (see in
Sect. IV), i.e.,
\begin{eqnarray}
 \label{pot}
V(x) = V_0 [\sin(2 \pi x) + c_1 \sin (4 \pi x) + c_2 \sin (6 \pi x)],
\end{eqnarray}
wherein $V_0$ normalizes the barrier height to unity, and the parameters
$c_1$ and $c_2$ characterize the spatial asymmetry.

\section{Fluctuation and rectification measures}

Throughout the following we focus on the asymptotic, periodic
 regime after effects due to the influence of initial
conditions and transient processes have quiet down.  Then, the main
statistical quantifiers of the driven stochastic process can be
described in terms of time- and ensemble-averages.  For a given
quantity $f(x(t))$ its time-homogeneous statistical properties are
obtained only in the long-time limit after transients have died out and
after both, the average over the temporal period of the
driving and  the corresponding ensemble-average are performed
\cite{jung1990}. In this asymptotic regime the
 time-independent (single-time) quantities are obtained
 by a double averaging procedure over both the noise and the
 period of driving; i.e.,

\begin{eqnarray}
 \label{aver}
\langle f \rangle = \lim_{t\to\infty} \frac{\omega}{2\pi}
\int_{t}^{t+2\pi/\omega} \prec f(x(s)) \succ \; ds,
\end{eqnarray}
wherein $\prec f(s) \succ$ indicates the average over the {\it noise realizations}
(ensemble-average).

The most salient transport quantity is the average, directed velocity
$\langle v \rangle$ of the driven Brownian particle.
Here $v(t)$ denotes the stochastic process $\dot x(t)$ in Eq. \ref{NLbw}.
Of equal
importance are, however, the fluctuations of $v(t)$ around its mean $\langle v \rangle$
in the long time regime, i.e., the variance
%
\begin{eqnarray}
 \label{sigma}
\sigma _{v}^2 = \langle v^2 \rangle - \langle v \rangle^2.
\end{eqnarray}
The Brownian motor moves with current values $v(t)$ that
range typically within
\begin{eqnarray}
 \label{v(t)}
v(t) \in \left[ \langle v \rangle -\sigma_v, \langle v \rangle
+\sigma_v\right].
\end{eqnarray}
If $\sigma_v > \langle v \rangle$, and even more so if $\sigma_v \gg
\langle v \rangle$, the Brownian motor can possibly move for some time
in the opposite direction of its average value $\langle v\rangle$.
The question thus arises: Is an efficient directed transport still
feasible?

To answer this challenge we shall introduce a measure for the
efficiency $\eta$ of the rectification of thermal noise,
a quantity directly related to the velocity fluctuations.
Here, we follow the
reasoning of Suzuki and Munakata \cite{Suzuki2003}, which yields a
nonvanishing rectification efficiency also in the absence of an external
bias.  This efficiency of rectification follows from an energy balance
of the underlying inertial Langevin dynamics. When specialized to our
situation, $\eta$ is given by the ratio of the dissipated power
$\gamma \langle v \rangle^2$ associated with the directed
motion of the motor against friction, and the input power from the
time-periodic forcing.  The result assumes the explicit form (see in
the Appendix)
\begin{eqnarray}
\label{effic1}
\eta & = & \frac{ \langle v \rangle^2} {|\langle v \rangle^2
+\sigma_v^2 - D_0|}\;, \nonumber \\ & = & \frac{\langle v \rangle^2}
{|\langle v^2 \rangle - D_0|}\;.
\end{eqnarray}

It thus follows that for a decreasing variance of the velocity
fluctuations, $\sigma_v^2$, the efficiency of the Brownian motor
{\it increases}. This is just what one would expect on naive grounds:
The transport of a Brownian motor can be optimized in regimes of a
large, directed average current which intrinsically does exhibit only
small fluctuations.  Moreover, in our study we found numerically that
$\langle v^2 \rangle > D_0$ holds true for any chosen set of
the simulation parameters.

\section{Fluctuation behavior of current in an inertial
rocked Brownian motor}

Deterministic inertial Brownian motors exhibit a complex dynamics
including chaotic regimes.  The application of noise then generally
destroys the complex fine structure of their phase space and tends to
smooth out their characteristic response function.

There are two classes of states of the driven system dynamics: the
locked states, in which the particle stays inside one potential well,
and the running states for which the particle runs over the potential
barriers. The first regime is characteristic for small driving
strengths. When the amplitude of the external force is made
sufficiently large we find that running states appear. These running
states can either be chaotic (diffusive) or regular.

\subsection{Numerical schemes}

We have carried out extensive numerical studies in order to identify
generic properties of the noise-induced transport.  Applying a small,
but finite noise strength, $D_0 >  0$, we have integrated the
Langevin equation (\ref{NLbw}) by employing the Euler method with a
time step of $h = 10^{-3}$. For the initial condition of the
coordinate $x(t)$ we used a uniform distribution over the
dimensionless period $L=1$ of the ratchet potential.  Likewise, the
(scaled) starting velocity has been chosen at random from a symmetric,
uniform distribution over the interval $[-1, 1]$.  All quantities of
interest were averaged over 250 different trajectories. Each single
trajectory evolved over $45 \times 10^3$ periods.
The transient regime
usually relaxed already long before 500 periods of the driven dynamics
had elapsed.
For the cases of very weak noise and weak driving we extended the
corresponding time-span to ensure that the transient dynamics has
quiet down completely.
In the limiting deterministic case, i.e. $D_0 = 0$, we
used the Runge-Kutta algorithm of order 5.  In this case, the averages
were calculated over $10^3$ differing trajectories, each
trajectory evolving over $10^3$ periods.

\begin{figure}[htbp]
\centerline{\includegraphics[angle=0,scale=.6]{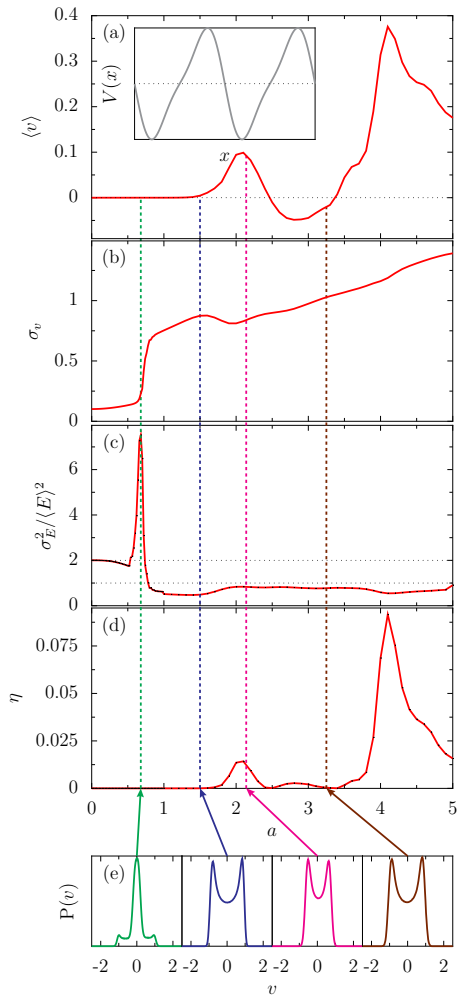}}
\caption{\label{figJA}(Color online)
  Fluctuation behavior of an inertial Brownian motor versus the
  driving strength $a$.  (a): averaged dimensionless velocity $\langle
  v\rangle$ of the inertial Brownian motor in Eq. (\ref{NLbw}); (b):
  variance of the corresponding velocity fluctuations $\sigma_v$; (c):
  fluctuations of the rescaled kinetic energy $\sigma_{E}^2 / \langle
  E \rangle^2$; (d): rectification efficiency in Eq. (\ref{effic1}).
  All quantities have been computed for the rescaled potential $V(x) =
  - V_0 [\sin(2 \pi x) + 0.25 \sin (4 \pi x)]$, where $V_0 \simeq
  0.454$ normalizes the barrier height to unity, see the inset in (a).
The force corresponding to this potential ranges from $-2.14$ to
$4.28$. The angular frequencies  at the well-bottom and at the
barrier-top, respectively, equal each other, reading $5.28$.
Bottom panel (e): velocity distributions for selected driving
amplitudes, i.e. $a$ = 0.68, 1.5, 2.14, 3.25. All these
distributions were normalized by setting their maximum to one.  The
remaining rescaled parameters read: friction $\gamma=0.5$, angular
driving frequency $\omega = 3.6$ and weak thermal noise of strength
$D_0=0.01$.  }
\end{figure}

\subsection{Numerical results: Current fluctuations versus driving
 strength}
We start out to study the role of fluctuations by varying the
amplitude $a$ of the sinusoidal driving force. In doing so, we
assume a relatively small temperature, so that the Brownian motor
dynamics is not far from a deterministic behavior as described in prior
works \cite{Jung1996, Mateos2000,Borromeo2002}.  The
average asymptotic long-time current velocity is shown in Fig. \ref{figJA}(a).
It reveals
that  for an amplitude $a\simeq 1.5$ the directed, {\it inertial} transport
sets in before the lower threshold of the ratchet force is reached.
It assumes a first local maximum
near the lower threshold of the
potential force near $a \simeq 2.1$.
Note that in the presence of small noise, the current is strictly
speaking never zero. For all practical purposes, however, we can
characterize the outcome of our Langevin simulation
as a deterministic, zero-current result. Below this threshold the system mainly
dwells in the locked state.  Upon closer inspection, we notice that in the
vicinity of $a\simeq 0.6$, the velocity
fluctuations $\sigma_v$ shown in Fig. \ref{figJA}(b) undergo a rapid
increase.  In Fig. \ref{figJA}(c) we also display the relative
fluctuations of the kinetic energy, $E=v^2/2$, (the re-scaled mass is
one); i.e.
\begin{eqnarray}
 \label{energy}
\frac{\sigma^2_{E}}{\langle E \rangle^2} = \frac{ \langle v^4 \rangle
- \langle v^2 \rangle^2} {\langle v^2 \rangle^2} \;\,.
\end{eqnarray}
Around this value of the driving amplitude, such a quantity
undergoes a giant enhancement.  Finally, we remark that for the equilibrium
Maxwell distribution we find $\sigma^2_{E}/\langle E \rangle^2
=2$, as expected.

Upon further increasing the amplitude of driving, $a>1.5$, the
Brownian motor generates for this set of parameters the desired,
directed transport behavior.  At the same time we observe that the
width of the weakly asymmetric, corresponding  distribution $P(v)$
slightly decreases, meaning
that the velocity fluctuations become smaller. The following
explanation thus applies: Because at $a<1.5$ escape jumps between the
neighboring wells are rare, i.e., the average directed current is very
small (note also the accompanying, very weak asymmetry in the velocity
distribution).  The input energy is pumped primarily into the kinetic
energy of the intra-well motion and eventually dissipated. As $a$ is
increased further, the Brownian motor mechanism starts to work, and
some part of energy contributes to the net motion of the particle.
Therefore, less energy remains available to drive intra-well
oscillations and consequently the distribution $P(v)$ shrinks, see in
Fig. \ref{figJA}(e)

Correspondingly, due to inertia, the mean velocity increases, reaching
a second maximum before the upper threshold value of the potential
force $a\simeq 4.28$.  Above this driving amplitude, the current
starts to decrease because of the weakening influence of the ratchet
potential at large rocking amplitudes.

The occurrence of multiple reversals for the directed current, as it
occurs in Fig. \ref{figJA}(a), is a known, interesting feature of
inertial Brownian motors.  Several prior studies did elucidate in
greater detail the corresponding mechanism at work \cite{Jung1996,
  Mateos2000, Barbi, Borromeo2002, Family}.  Here, we take instead a
closer look at the current fluctuations.  We observe that for the
chosen set of parameters the maximal stationary velocity in Fig.
\ref{figJA}(a) does not exceed the value 0.4. In contrast, its
fluctuations keep growing as the driving amplitude rises. At large
driving, the particle no longer feels the potential and undergoes a
rocked, free Brownian motion with the velocity fluctuations growing
proportional to $a$, cf. Fig. \ref{figJA}(b).

On the other hand, the relative fluctuations of the kinetic energy do
saturate, see Fig. \ref{figJA}(c).  These are suppressed to values
near $1$, which lies below the equilibrium value of $2$.  Within this
directed transport regime, the efficiency (\ref{effic1}) remains
rather small, cf Fig. \ref{figJA}(d). Such small rectification
efficiency is the rule for this driven inertial Brownian motor.

Let us next inspect the {\it current probability distribution} $P(v)$.
All $P(v)$ curves reported in the following have been normalized so
that their maximum (i.e., their highest peak) is set to a fixed, unit
value.  Only then we can detect the details in their shape upon
varying the corresponding parameter such as the driving amplitude $a$
or the noise strength $D_0$.  These probabilities look rather
symmetric; however, a finite ratchet velocity requires a certain
amount of asymmetry either in the location and/or the width of the
velocity peaks.  Here, the current results mainly due to a slight
shift of the maxima location.

The most peculiar feature of the current distributions shown in Fig.
\ref{figJA}(e) is the emergence for $a>0.6$ of two additional
side-peaks centered near $v=\pm 1$, which eventually dominate $P(v)$
at larger driving amplitudes.  Of course, for zero drive $P(v)$ boils
down to a single-peaked Maxwell distribution, strictly symmetric
around $v=0$.  To investigate the onset of these two side-peaks for
$a$ values corresponding to vanishingly small currents, we set $P(v)=
q P_0(v-1) + q P_0(v+1) + (1-2 q) P_0(v)$, where $q$ varies from $0$
to $1$. If $P_0$ is taken to be a symmetric Gaussian function, then
the kinetic energy fluctuations can be evaluated; $\sigma_E$ exhibits
the behavior shown in Fig. \ref{figJA}(c).

What it the origin of those three peaks in the distribution $P(v)$?
Our first conjecture to connect it with the 'running' solutions turned
out to be incorrect.  This is so, because for $a\lesssim 1$ the
particle rarely leaves the confining potential well and thus cannot
significantly contribute to the side peaks of the distribution
function.  We further checked the outcome for the velocity
distribution when reflecting barriers were placed at the maxima of the
potential. Under such constraints, the three-peak-structure is
recovered as well.  Moreover, the sinusoidally driven damped particle
in a harmonic potential can exhibit both, a singly-peaked as well a
doubly-peaked averaged velocity distribution, see in Ref.
\cite{jung1990}.  However, for the parabolic potential that fits best
the wells of our ratchet potential around its minima, we found a
single peaked $P(v)$.

We therefore do conclude that the
characteristic behavior for the additional side-peaks is rooted in the
nonlinear, anharmonic character of the corresponding well of the
periodic asymmetric ratchet profile.

\begin{figure}[htbp]
\centerline{\includegraphics[angle=0,scale=.62]{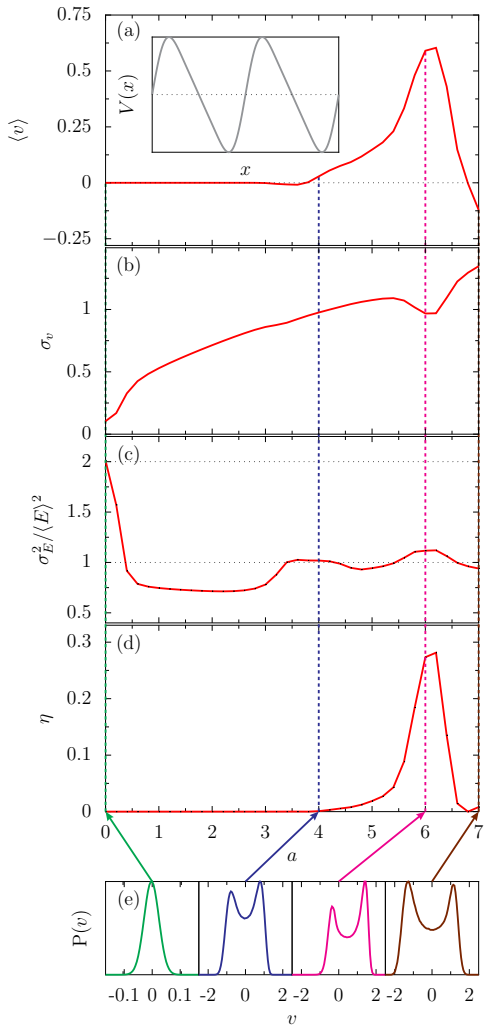}}
\caption{\label{figJA2}(Color online)
  Tailoring the shape of the potential. (a): averaged dimensionless
  velocity $\langle v\rangle$ of the inertial Brownian motor under
  nonadiabatic driving conditions; (b): corresponding velocity
  fluctuations $\sigma_v$; (c): fluctuations of the rescaled kinetic
  energy $\sigma_{E}^2 / \langle E \rangle^2$; (d): efficiency. All
  quantities are plotted versus the external driving amplitude $a$ for
  the asymmetric ratchet potential $V(x) = V_0 [\sin(2 \pi x) + 0.245
  \sin (4 \pi x) + 0.04 \sin (6 \pi x)]$, where $V_0\simeq 0.461$
  normalizes the barrier height to unity (see inset in (a)).  The
  forces stemming from such a potential range between $-4.67$ and
  $1.83$. The two angular frequencies  at the well-bottom and at the
  barrier-top are  the same, reading
   $5.34$.
The corresponding velocity distributions $P(v)$ are displayed in panel (e)
for the indicated driving amplitudes, i.e., $a$ = $0$, $4$, $6$,
$7$. All $P(v)$ curves have been normalized by setting their maximum to one.
The remaining parameters are: $\gamma=0.9$, $\omega =
4.9$ and $D_0=0.01$.  }
\end{figure}

\subsection{Numerical results: role of the shape of the underlying ratchet potential}
We next study the influence of the ratchet profile (or force) for the
mean velocity and the corresponding fluctuation behavior of the
directed current. We use a stylized potential shape composed of three
spatial higher harmonics with $c_2 \neq 0$, see in (\ref{pot}). Note
that this potential shape possesses an opposite polarity as compared
with the ratchet potential depicted in the inset of Fig.
\ref{figJA}(a).  Put differently, the natural direction of the
Brownian motor motion is to the left, in the direction of the weaker
slope, see Fig. \ref{figJA2}(a), inset.

For  small driving amplitudes, this inertial motor predominantly
dwells in a potential well. The directed current is very small and
negative (directed towards the left).
If the driving amplitude exceeds the upper
threshold amplitude of the ratchet force at $a\simeq 4.66$,
the motor starts to move more regularly, reaching an
extremal speed near $a \simeq 6$. Interestingly enough, the motor moves now towards
 {\it positive} $x$-values. The resulting
transport velocity thus cannot be easily predicted {\it a priori}
in this nonadiabatic driving regime.  This is
a benchmark feature of these inertial Brownian motors where the coupling
between the deterministic driving force, $F(t)=a\cos(\omega t)$, and
the resulting motion of the driven Brownian particle are coupled
loosely, only. Except for a narrow regime of sizable values, cf. Fig. \ref{figJA2}(a),
the emerging average velocities are typically very small, yielding
no corresponding  efficient rectification. Therefore, the efficiency of
this more complex Brownian motor mimics again closely the behavior
of the average motor velocity,  see panel (d) in Fig. \ref{figJA2}.

The velocity fluctuations  exhibit a similar behavior as in the case
of Fig. \ref{figJA} discussed above.
The variance grows nearly linearly with increasing driving strength.
Typically, the average velocities are small and the fluctuation behavior is
similar to the behavior discussed above for the first ratchet potential.
Interestingly enough, however, above threshold driving
induces   a  distinct peak behavior for the velocity which
is accompanied by a corresponding dip in the
behavior for the velocity fluctuations. This dip in the variance then gives rise to
a window with appreciable efficiency,
cf. Fig. \ref{figJA2}(d).

The behavior for the velocity distributions is again generic: Small average velocities
exhibit  nearly symmetric velocity distributions, see Fig. \ref{figJA2}(e).
Only for the nonadiabatic peak behavior
of the mean velocity does one identify also an appreciable asymmetry for the velocity distribution.

\subsection{Numerical results: Current fluctuations versus noise strength $D_0$}
\begin{figure}[htbp]
\centerline{\includegraphics[angle=0,scale=.69]{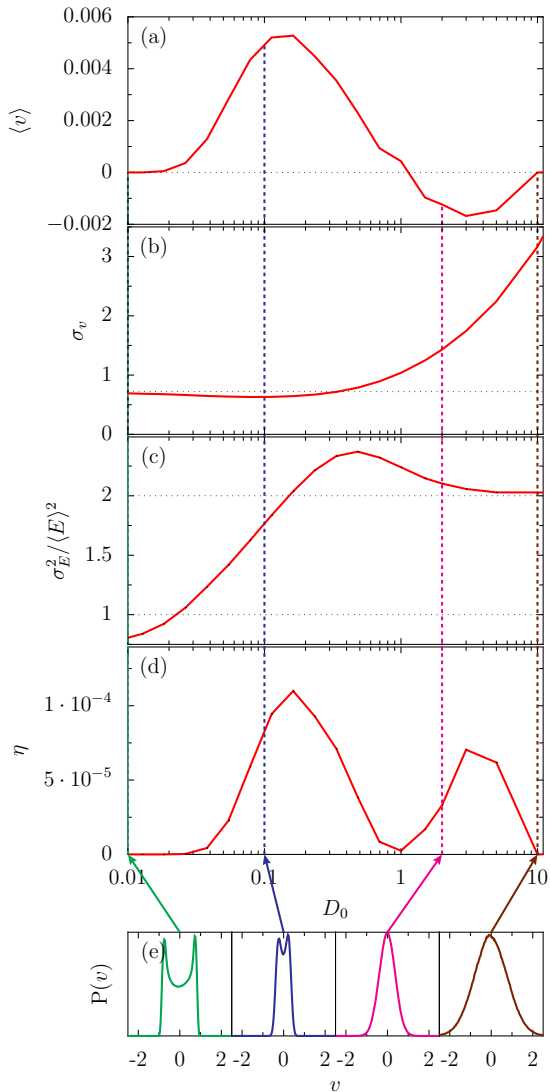}}
\caption{\label{figJKT} (Color online)
Fluctuation behavior of an inertial Brownian motor versus the noise
strength $D_0$.  (a): averaged dimensionless velocity
$\langle v\rangle$; (b): the corresponding velocity variance
$\sigma_v$;  (c): fluctuations of the rescaled kinetic energy
$\sigma_{E}^2 / \langle E \rangle^2$; (d): the corresponding
efficiency.  Numerical results obtained for the same ratchet potential as in
Fig. \ref{figJA}.  In the bottom panel (e), the velocity
distribution $P(v)$ is shown for $D_0$ = 0.01, 0.1,
2, 10. All $P(v)$ curves are normalized
as in Fig. \ref{figJA}. The remaining rescaled parameters are: $\gamma=0.5$,
$\omega = 3.6$ and $a=0.8$.  }
\end{figure}

%
In Fig. \ref{figJKT}, we numerically investigate the directed
transport versus the temperature $D_0$.  We have chosen a
sub-threshold driving strength for which the thermal noise plays a
constructive role \cite{SR} by inducing noise activated jumps
across the potential barriers. We set $a=0.8$ and the other parameters
remain the same as in Fig. \ref{figJA}. Then, we find a characteristic
velocity reversal near the dimensionless temperature $D_0=kT/\Delta V
\simeq1$. There, the thermal energy compares with the activation
energy over the barrier height of the ratchet potential.  A subsequent
increase of temperature causes a diminishing role of the asymmetric
ratchet potential and, consequently, the directed transport degrades.

Moreover, the time-averaged velocity distribution approaches the
equilibrium velocity distribution \cite{Schreier1998}.  It is
remarkable that within a certain range of temperatures the
fluctuations of the kinetic energy exceed the relevant value for the Maxwellian
equilibrium distribution, cf. Fig. \ref{figJKT}(c).  A shallow, local
minimum occurs for the velocity fluctuations where the average current
itself is maximal. These fluctuations are, however, notably three
orders of magnitude {\it larger} than the small-valued, directed
current. Not surprisingly, the rectification efficiency shown in
Fig. \ref{figJKT}(d) is quite small.  Again, the Brownian motor is not
operating optimally.

\section{Tailoring rectification efficiency}

Thus far, changing the ratchet profiles
did not lead to a large enhancement of the
rectification efficiency.  What is needed in achieving a large
rectification efficiency is a sizable Brownian motor current which is
accompanied by small current fluctuations only, see
Eq. (\ref{effic1}). This scenario seemingly implies that the directed
current should proceed in a persistent manner with very few,
occasional back-turns only. This in turn causes small fluctuations in
the velocity and, additionally, provides a dominating  asymmetry
of the velocity distribution.

Such a behavior can be realized by a combined tailoring of the
asymmetry of the ratchet potential together with the use of
appropriate driving conditions.  In the quest for achieving such a
favorable situation we use the three-harmonics ratchet potential
plotted in the inset of Fig. \ref{figJA2}(a). Our hope is that upon
minimizing the noise further we can achieve a substantial improvement
of the efficiency.

\begin{figure}[htbp]
\centerline{\includegraphics[angle=0,scale=.7]{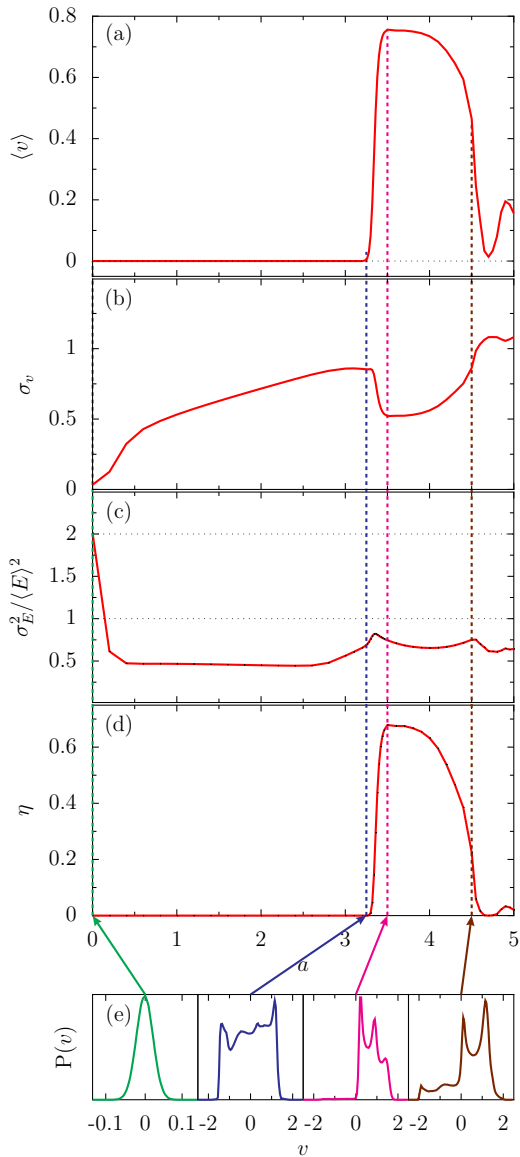}}
\caption{\label{figJGAMMA}(Color online)
Tailoring the shape of the potential. (a):
averaged dimensionless velocity $\langle v\rangle$ of the inertial
Brownian motor under nonadiabatic driving conditions;
(b): corresponding velocity fluctuations
$\sigma_v$; (c): fluctuations of the rescaled kinetic energy
$\sigma_{E}^2 / \langle E \rangle^2$; (d): efficiency. All quantities are
plotted versus the driving amplitude $a$ for the
asymmetric ratchet potential of Fig. \ref{figJA2}.
%
The corresponding velocity distributions $P(v)$ are shown in panel (e)
for selected driving
amplitudes, i.e., $a$ = $0$, $3.25$, $3.5$,
$4.5$. All $P(v)$ curves are normalized as in Fig. \ref{figJA2}.
The remaining parameters are: $\gamma=0.9$, $\omega =
4.9$ and $D_0=0.001$.  }
\end{figure}

At {\it very weak noise} and large, nonadiabatic rocking frequencies,
this inertial Brownian motor starts moving efficiently near the upper
threshold of the ratchet force $a \simeq 4.66$, see
Fig.\ref{figJGAMMA} (a).  Because the directed velocity becomes
maximal and simultaneously its fluctuations are locally minimal, see
in Fig. \ref{figJGAMMA}(b), we indeed find the desired enhancement of
the rectification efficiency, see Fig. \ref{figJGAMMA}(d).  The
fluctuations of the kinetic energy grow slightly; nevertheless, these
are still strongly suppressed in comparison to the equilibrium value
$2$.

We have studied several other ratchet potentials by varying the
parameters $c_1$ and $c_2$ in Eq.(\ref{pot}) and still found regimes
where the inertial ratchet works with a high efficiency (not shown).
In all these cases we found that the velocity distribution has a
support concentrated mainly on one of the semi-axes. Strongly
asymmetric velocity distributions are depicted with Fig.
\ref{figJGAMMA}(e). In contrast, with the mode $c_2$ set zero (see in
inset in \ref{figJA}) we could not identify such an optimal regime for
rectification of noise.  The shape of these distributions just
corroborates the fact that large rectification efficiencies are the
result of persistent, (uni)-directional Brownian motor motion,
accompanied by a strong asymmetry of the current statistics.

\section{CONCLUSION}

With this work we have elucidated, directed Brownian motor transport
in rocked rachet potentials in the presence of inertia and thermal
noise.  We focussed on several parameter regimes and studied by
numerical means the operation of this massive ratchet machine. In
particular, we investigated the variation of the average current
versus driving amplitude $a$ and the temperature strength $D_0$. Our
main objective has been  the behavior of the accompanying current
fluctuations as a function of these transport parameters.  These
fluctuations crucially impact the rectification behavior, as measured
by the rectification efficiency in Eq. (\ref{effic1}).

Typically, the current values and the corresponding velocity
fluctuations are so, that no appreciable rectification emerges in
these inertial, rocked Brownian motors. There exist, however, tailored
regimes of rachet profiles and driving parameters for which an
enhancement of rectification and optimal transport do occur. These
regimes are marked by a large Brownian motion transport with few
back-turns only. This in turn implies a narrow, asymmetric velocity
statistics with dominantly, one-sided support of either positive- or
negative-valued velocities.

These novel findings for the fluctuation statistics of Brownian motor
velocities can be put to use in diverse technological devices that
pump and separate efficiently and reliably Brownian particles in
correspondent physical \cite{BM} and biological Brownian motor systems
\cite{molecular}. Moreover, the results derived herein for such driven
inertial Brownian motors can be applied as well to the phenomenon of
Stochastic Resonance \cite{SR} in corresponding underdamped regimes.

\section*{Appendix}
%
%
With this appendix we present the derivation of the expression
(\ref{effic1}) for the rectification efficiency.
We combine the arguments in \cite{Derenyi1999,Oster,Suzuki2003}
and establish the efficiency $\eta$ as
\begin{eqnarray}
\label{etaAPP}
\eta = {A \over |P_{in}|}
\end{eqnarray}
In the denominator, $P_{in}$ denotes the rate of the energy input to
the system.  There is no overall consensus on the numerator $A$
\cite{efficiency,Derenyi1999,Oster,Suzuki2003}.  If $A$ denotes the
rate of work done on the fluid by the Brownian motor motion, then the
corresponding efficiency $\eta$ is not an appropriate measure because
$A \sim \langle v^2 \rangle$. This quantity can be relatively large
even if there occurs no transport of the motor, i.e. even if $\langle
v \rangle = 0$ !  More suitable information on the efficiency of the
transport is gained when \cite{Derenyi1999,Oster} $A \sim \langle v
\rangle$. Following the reasoning in
\cite{Derenyi1999,Oster,Suzuki2003}, we use for the output power the
average friction force times the average velocity, i.e. $A = \langle
\gamma v \rangle \langle v \rangle $.  To calculate $P_{in}$, let us
recast (\ref{NLbw}) into the form
\begin{eqnarray}
\label{LAPP}
dx &=& v dt\;,\\
dv &=& - \Big( \gamma v + V'(x,t) \Big) dt + \sqrt{2 \gamma D_0} dW(t)\;,
\end{eqnarray}
where $V(x,t) = V(x) - a x \cos(\omega t)$ and $W(t)$
is the Wiener process
$( \langle W(t) \rangle = 0, \langle W^2(t) \rangle = t)$.

Now, we  evaluate the  ensemble and temporal averages of the re-scaled kinetic
energy $G(v) = v^2/2, \ v=v(t)$.
To this aim, first we apply Ito´s differential calculus to the function
$G(v)$ to obtain
\begin{eqnarray}
\label{ItoAPP}
d\Big( v^2/2 \Big) = - \Big( \gamma v^2 + v V'(x,t) - \gamma D_0 \Big) dt\\\nonumber
+ \sqrt{2 \gamma D_0} v dW(t).
\end{eqnarray}
The ensemble average (i.e. the average over all realization of the
Wiener process denoted by $\prec ... \succ$)
for the rate of change of  the kinetic energy  results in
\begin{eqnarray}
\label{ItoAPP}
{d\over dt} \prec v^2/2 \succ &=& - \Big[ \gamma \prec  v^2 \succ +
\prec  v V'(x) \succ  \nonumber \\
& & - \prec  v \;a \cos(\omega t) \succ - \gamma D_0 \Big]\;,
\nonumber\\
\end{eqnarray}
where we exploited the (Ito)-martingale property (for the part
containing the Wiener process).  Next, we average over the temporal
period as in (\ref{aver}) (periodic time-dependence of asymptotic
probability).  In doing so, we evaluate
\begin{equation}
\langle {d\over dt}  v^2 \rangle =
\prec v^2(t+2\pi/\omega)\succ-\prec v^2(t)\succ =0 \;.
\end{equation}
Likewise,  for the contribution
\begin{equation}
\langle v V'(x) \rangle=
\prec V\left(x(t+2\pi/\omega)\right)\succ-\prec V\left(x(t)\right)\succ = 0  \;.
\end{equation}
Consequently, we obtain
\begin{eqnarray}
\label{zero}
0= -\gamma \Big[ \langle v^2 \rangle - D_0\Big] + P_{in}\;,
\end{eqnarray}
where the combined average $P_{in}= \langle v(t) a \cos\omega t)
\rangle$ is the input energy to the system per unit time.  Thus, upon
combining (\ref{etaAPP}) and (\ref{zero}) the relation in
(\ref{effic1}) emerges. We also emphasize here, that our scheme for
the efficiency of rectification at zero bias is {\it independent} of
the transport friction-coefficient $\gamma$.  This feature is in
agreement with the corresponding result by Suzuki and Munakata
\cite{Suzuki2003}.

\section*{Acknowledgment}

The authors gratefully acknowledge financial support by the DAAD-KBN
(German-Polish project {\it Stochastic Complexity}) (PH and J{\L}),
the Foundation for Polish
Science [Fundacja na Rzecz Nauki Polskiej] (PH),
the ESF (Program {\it Stochastic Dynamics}), the Deutsche
Forschungsgemeinschaft via grant HA 1517/13-4, the Graduiertenkolleg
386 (P.H.,P.T.) and the collaborative
research grant SFB 486.

\end{document}